\documentclass[a4paper,12pt,oneside]{amsart}
\usepackage{graphicx}
\usepackage{amssymb}
\usepackage{paralist}

\usepackage{amscd}
\usepackage{dsfont}
\usepackage{bm}

\newtheorem{theorem}{Theorem}[section]

\newtheorem{conj}[theorem]{Conjecture}

\newtheorem{rems}[theorem]{Remarks}

\theoremstyle{definition}

\theoremstyle{remark}

\numberwithin{equation}{section}

\newcommand{\diag}[1]{\textrm{diag}\set{#1}}
\newcommand{\abs}[1]{\left\vert#1\right\vert}
\newcommand{\set}[1]{\left\{#1\right\}}

\def \1{\mathds{1}}
\def \0{\mathsf{0}}

\def\Cl{C}
\def\Z{ \mathcal{Z}}
\def\N{ \mathcal{N}}
\def\Q{ {\mathcal{Q}}}
\def\C{ {\mathcal{C}}}

\def\Kt{ K_{\scriptscriptstyle\mathrm{TrUE}}}
\def\Kk{ \mathcal{K}}
\def\Kkt{ \mathcal{K}_{\scriptscriptstyle\mathrm{TrUE}}}
\def\rhot{ \rho_{\scriptscriptstyle\mathrm{TrUE}}}

\def\trace{ {\mathrm{Tr}}}

\begin{document}

\title[]{Clustering of periodic orbits and ensembles of truncated unitary matrices}

\author{Boris Gutkin${}^\dag$, Vladimir Osipov${}^*$}
\address{
${}^\dag$ Faculty of Physics, University Duisburg-Essen, Lotharstr. 1, 47048 Duisburg, Germany;
\newline ${}^*$ Institute of Theoretical Physics, Cologne University
Zülpicher Str. 77, 50937 Cologne, Germany.
}
\email{boris.gutkin@uni-duisburg-essen.de}
\begin{abstract}
Periodic orbits in chaotic systems form clusters, whose elements traverse approximately the same points of the phase space. 
The distribution of cluster sizes depends on the length $n$ of orbits and the parameter $p$ which controls closeness of orbits actions.  We show that  counting of cluster sizes in the baker's map can be turned into a spectral problem for an ensemble of truncated unitary matrices. Based on the conjecture of the universality for the eigenvalues distribution at the spectral edge  of these ensembles, we obtain asymptotics of the second moment of cluster distribution  in a regime  where both  $n$ and $p$ tend to infinity. This  result  allows us to estimate the average cluster size as a function of the  number of encounters in periodic orbits.  

\end{abstract}

\maketitle

\section{Introduction}

\subsection{Motivation.} Periodic orbits have been at core of much of the mathematical work on the theory of the classical
and quantum  dynamical systems.
One of the primary  motivations to consider periodic orbits in Quantum Chaos  stems  from their  role as a bridge between two theories. In the semiclassical limit spectral density of chaotic Hamiltonian systems can be expressed  by the Gutzwiller trace formula through the sum over periodic orbits $\gamma$, where each term is determined by the  stability and  the action of $\gamma$ \cite{haake}. Straightforwardly,  this implies that spectral correlations between energy levels of quantum chaotic systems must be related to   correlations between actions of periodic orbits. This remarkable connection has been  fruitfully employed in a last decade to deduce universal features of quantum spectrum in chaotic systems described by Random Matrix Theory \cite{sr,haake1}.

 On a heuristic level, the action correlation mechanism can be understood in the following way.  A generic long periodic orbit in a chaotic system has a number of \textit{encounters}   where it  closely approaches itself in the phase space. It can be shown that for any such orbit  there exist partner orbits  following approximately the same path but  in a different time-order. The  switches in  the direction of motion occur at the encounters, see fig.~\ref{graphs}a. These periodic orbits form
   families, or \textit{clusters}  as we  refer to them hereafter.  All members of the  cluster  traverse through approximately the same points of the configuration space and, therefore, have close actions. Moreover, if the time reversal symmetry is broken  the orbits from the same cluster   must be close to each other also  in the phase space of the system.

The clustering phenomenon can be  rigorously described using symbolic dynamics. Assuming that dynamical system allows a finite Markov partition any periodic orbit can be represented as 
a cyclic sequence of symbols from certain alphabet \cite{licht}:
\[x=\overline{x_1x_2 \dots x_n}. \]
To simplify the treatment,  we will assume in this paper the simplest possible symbolic dynamics -- alphabet of two symbols $x_i\in\{0,1\}$  with any sequence  allowed by the trivial grammar rules. Such symbolic dynamics appears for instance in baker's map \cite{backer}. 
 Two periodic orbits from  the same cluster induce two sequences $x$, $x'$  with the  property, that each subsequence of $p$ consecutive symbols in  $x$ appears the same number of times  in $x'$, and vice versa. Accordingly, all periodic orbits can be arranged into a number of clusters. It is straightforward to see by the above definition that two periodic orbits from the same cluster are separated by metric distances of the order $2^{-p}$ in the phase space \cite{go1}.  Consequently,  the  number $p$  controls  the differences  between  their actions.

In chaotic systems the number of periodic orbits   grows  exponentially with their length. Because of this,   there are in general many periodic orbits with close actions which  belong to different clusters. Nevertheless, clustering phenomenon plays  very important role in Quantum Chaos.  Due to their topological nature  the action correlations between orbits within the same cluster are rigid. Namely, periodic orbits from the same clusters keep their action differences small also under small perturbation of the system. It can be expected that  the only systematic contribution into spectral correlations comes from such orbits.   
Taking this into account it seems to be a natural and important  question to ask what is the distribution of cluster sizes  for given $p$ and $n$. Note that in a nutshell this is a purely combinatorial problem, as we need to count the number of sequences of length $n$ satisfying certain constraint.

\subsection{Formulation of the problem and main results.}  The  cluster distribution can be studied through  $k$-th moments:
\[\Z_k(n)=\sum_{ \C} |\C|^k,\] where the sum runs over all clusters and $|\C|$ is the number of periodic orbits in the cluster $\C$.
It worth noting that the first momentum $\Z_1(n)=\sum_{\C} |\C|$ provides just the total number of periodic orbits of the period $n$, whose asymptotics for the baker's map is given by $2^n/n$.
In this paper the  consideration will be restricted to the  second  moment $\Z_2(n)$. The ratio $\langle|\C|\rangle:=\Z_2(n)/\Z_1(n)$ then can be used as a measure of the average cluster size.

As can be easily understood the asymptotic behavior of $\Z_2(n)$ for long orbits crucially depends on the interplay between parameters $p$ and $n$.
 In  our previous work \cite{go1}  and related studies \cite{sharp1,sharp2} the long trajectory limit $n\to\infty$, with $p$ being fixed was considered. Note however, that in a proper semiclassical limit the  parameters  $n, p$ are related to each other.   
  Indeed,  the spectral correlations on the scale of mean level spacing $\Delta$ are determined by periodic  orbits with the periods   of the  order of the \textit{Heisenberg time} $\hbar\Delta^{-1}$.  For  two-dimensional systems this time scale is proportional to the inverse Planck's constant $\hbar^{-1}$. On the other hand, the action differences between correlating  orbits must be of the   order of Planck's constant.  Since  action differences and  periods of orbits  are controlled by $p$ and $n$ respectively, in a semiclassical  regime both parameters must tend to infinity.   
The main  focus of the present paper is on the  limit: $n=t\sqrt{N}$,  where $t$ is a fixed parameter and $N=2^{p-1} \to \infty$.   Speaking informally, this   regime  can be interpreted as  the limit where the  average number of encounters in periodic orbits is fixed.
  This is so, since for long periodic orbits with $n\ll N$ the number of encounters is proportional to 
$n^2/N$.
Note that this regime is  essentially   different from the aforementioned  semiclassical limit   needed for the evaluation of spectral correlation on the scale of mean level spacing. In the last case $N\sim n$ and the number of encounters in periodic orbits tends to infinity.

The central idea of our approach is to represent  $\Z_2(n)$ through the  spectral form factor of an ensemble of truncated $2N\times 2N$  unitary matrices. As we show, the relevant spectral information necessary for evaluation of $\Z_2(n)$ depends on the relationship between $n$ and $N$.  In the semiclassical regime  $N\sim n$ the asymptotic form of $\Z_2(n)$ is determined solely by the distribution of the largest eigenvalue in the ensemble of random matrices. On the contrary, in the regime $\sqrt{N}\sim n$  the relevant information  arrives from the edge of the spectrum, where both eigenvalues distribution and their correlations are of importance. Remarkably, the density and correlations of eigenvalues at the $1/\sqrt{N}$ vicinity of the spectral  edge exhibit universal properties. This observation allows us to obtain  asymptotics of $\Z_2(n)$ and estimate ``average`` cluster size  $\langle|\C|\rangle$ analytically. \\ 

The paper is structured as follows: In  Sec.~2  we relate function $\Z_2(n)$ to the spectral form factor of an ensemble
of truncated  $2N\times 2N$ unitary matrices. We then analyze the spectrum of these matrices in Sec.~3.  We demonstrate numerically that at the  edge of the spectrum a universality  of eigenvalue distribution holds. This universal behavior is then analyzed in Sec.~4 using  circular ensemble of   truncated  unitary (CUE) matrices  with the standard invariant measure.    For these ensembles we  analytically  derive asymptotics of the spectral form factor in the regime $\sqrt{N}\sim n$. In  Sec.~5 we exploit these results  in order to obtain    asymptotics of $\Z_2(n)$ and consider implications for the average size of clusters.
 Discussion of different regimes along with open questions for further research are put
in Sec.~6.

\section{Cluster counting as  spectral problem}
As a first step, we  show below that the problem of finding   cluster sizes  is equivalent to the one of counting closed paths of the same length on a special class of graphs  $G_p$. Specifically, for the symbolic dynamics of backer's map  $G_p$ happen to be  famous  \textit{de Bruijn graphs}. 
 
\subsection{Clusters of periodic orbits on graphs}
\begin{figure}[htb]
\begin{center}{
\includegraphics[height=5.5cm]{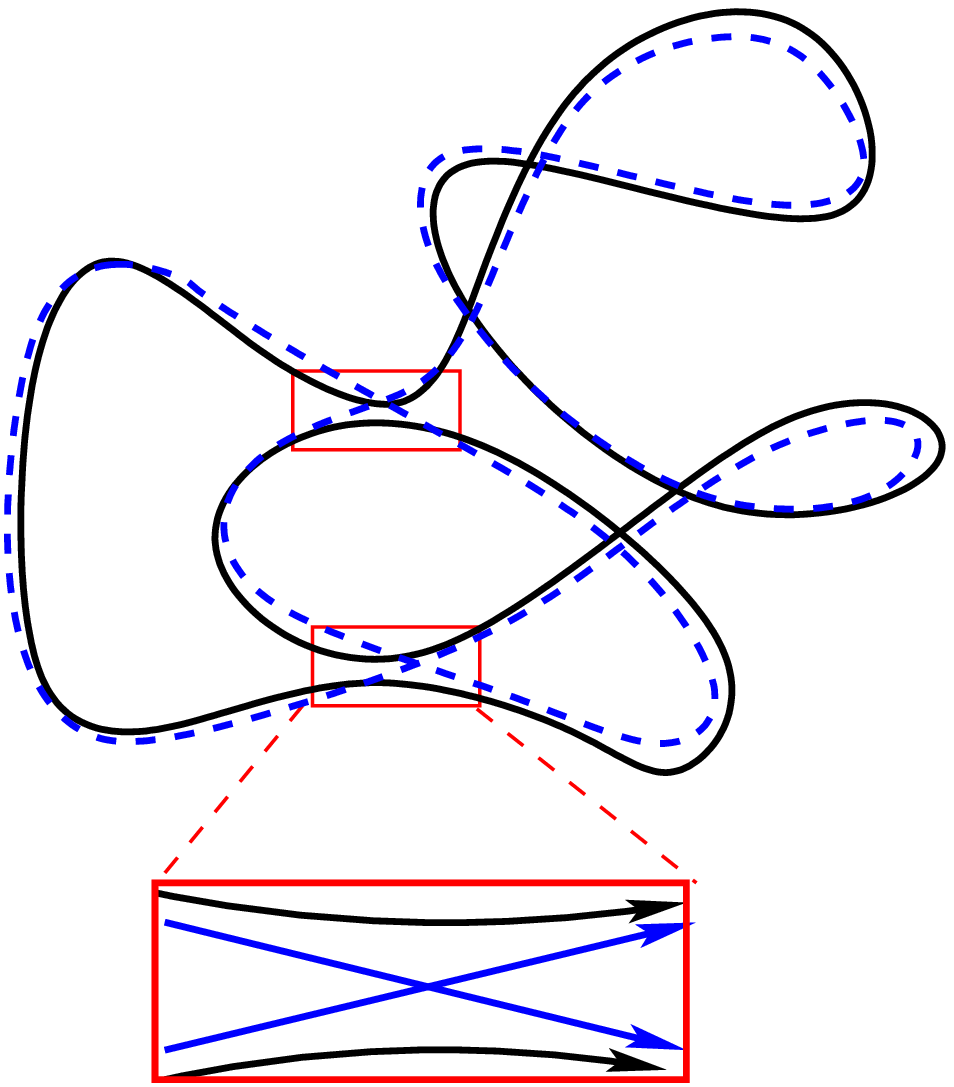}a) \hskip 1.2cm \includegraphics[height=5.5cm]{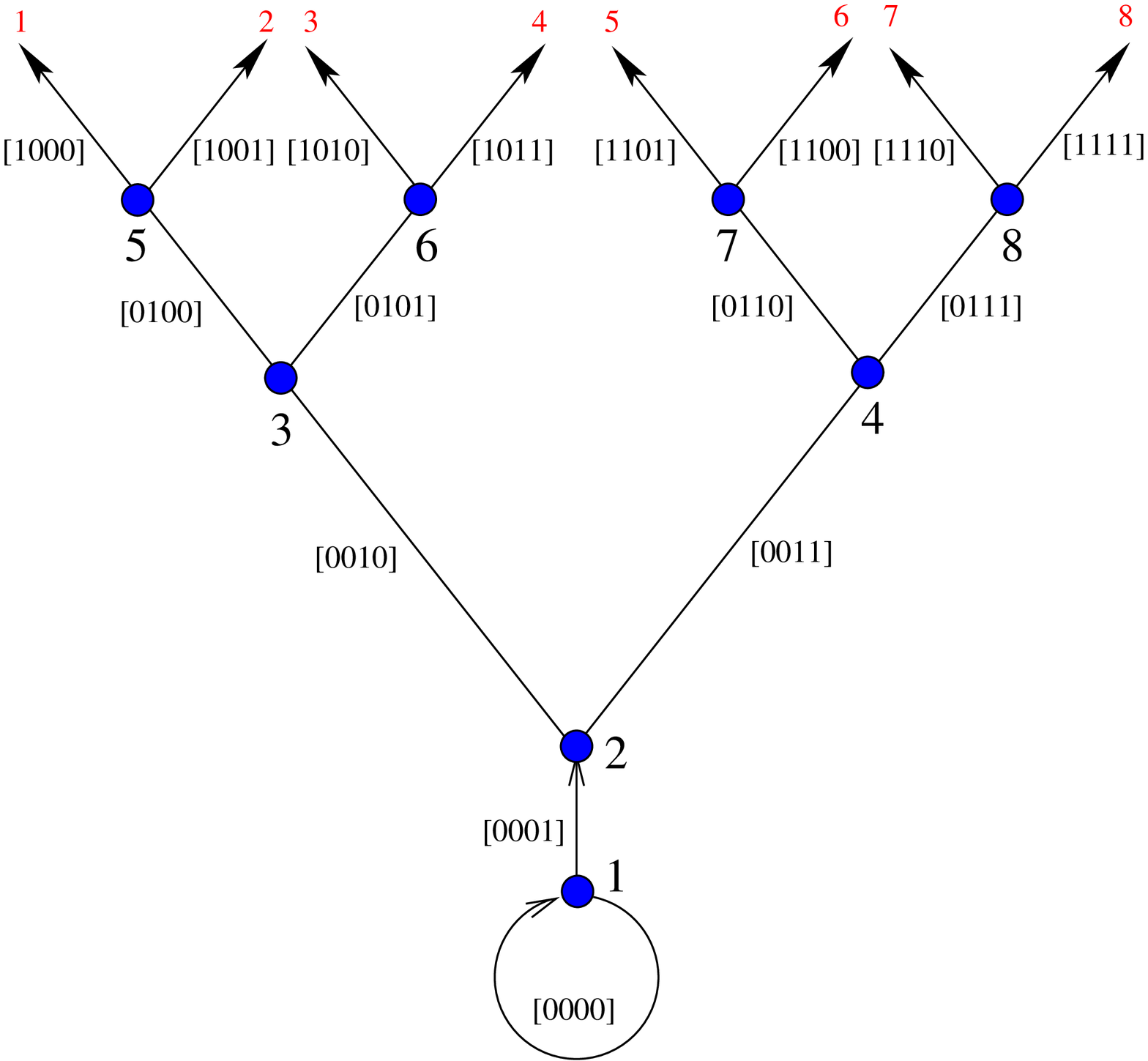}b)
}\end{center}
\caption{ \small{On the left is shown a caricature of two periodic orbits  belonging to the same cluster. 
Apart from the encounters (red rectangular boxes) two orbits closely follow  each other in the phase space. At the encounters orbits   switch the directions,  as shown in the inset.   On the right is the graph $G_p$  for  $p=4$. Each sequence $[a_1 a_2 a_3 a_4]$, $a_i\in\{0,1\}$   encodes edges of the graph.}  }\label{graphs}
\end{figure}

  Let $X_s$ be  the
set of all possible sequences of zeroes and ones having the length $s$. 
The graph $G_p$ is constructed in the following way. First, with each sequence $[a_1a_2\dots a_{p-1}]\in X_{p-1}$ we associate a vertex of  $G_p$. 
Second, a sequence $a=[a_1a_2\dots a_p]$, $a\in X_p$  defines the directed  edge $e_a$ of $G_p$ which connects vertex $[a_1a_2\dots a_{p-1}]$ to the vertex $[a_2a_3\dots a_p]$.
Thus the total number of edges is $2^p$ and each vertex has two incoming and two outcoming edges, see fig.~\ref{graphs}b. It  is worth mentioning that the resulting graphs   $G_p$ are actually well  known in the literature. These  are so-called  de Bruijn graphs, first introduced in \cite{Bruijn}.

It is straightforward to see that any closed path $\gamma_x$ on the graph  $G_p$  passing through $n$ edges can be uniquely  represented by a cyclic sequence $x=\overline{x_1x_2\dots x_n}$. By such identification each  edge of  $G_p$ traversed by $\gamma_x$ corresponds to a certain segment  $[x_{i\!\!\!\mod n} x_{i+1\!\!\!\mod n}\dots x_{p-1+i\!\!\!\mod n}]$ of the sequence $x$. 
Let $\bm n (x) = \{n_a, a\in X_p\}$ denote a set of integers, such that $n_a$ is the number of times  the path $\gamma_x$   passes through the edge $a$. Correspondingly $|\bm n (x)|=\sum n_a=n$ is the length of the path.  Then two sequences $x,y\in X_n$, belong to the same cluster  if and only if   $\bm n (x) = \bm n (y)$. Equivalently, $x,y$ are in the same cluster if $\gamma_x$ and $\gamma_y$ traverse  edges of $G_p$  the same number of times (which can be also zero). As a result, each cluster  of  periodic orbits can be unambiguously labeled by the vector of integers $\bm n=\{n_a, a\in X_p\}$. Keeping this in mind we introduce notation  $\C_{\bm n}$ for the cluster of periodic orbits corresponding to   vector $\bm n$ and aim to find  asymptotics of:
\begin{equation}
 \Z_2(n)=\sum_{ |\bm n|=n} |\C_{\bm n}|^2.
\end{equation}

\begin{rems}\label{remark} 
By attaching  incommensurate ``lengths'' to  edges of the graph $G_p$ one can turn it into a metric graph. Then each  $\C_{\bm n}$ 
can  be seen as a cluster of closed paths  with equal lengths. Accordingly, counting of cluster sizes is equivalent to counting of degeneracies in the length spectrum of the corresponding metric graph. The last problem has been studied in \cite{uzy2,sharp2,tanner,Berkolajko} for some classes of metric graphs. 
\end{rems}

Rather then consider  $\Z_2(n)$ directly it turns out to be convenient to introduce a slightly different function. Recall that each periodic orbit is uniquely identified by a sequence of symbols up to a cyclic shift. This means all $n$ sequences:
\[ [x_{(i\!\!\!\mod n)}x_{(i+1\!\!\!\mod n)} \dots x_{(i+n-1\!\!\!\mod n)}]\in X_n, \qquad i=1,\dots n\]
correspond to one and the same periodic orbit $\gamma_x$.
For  {\it prime} periodic orbits, which have no repetitions all these sequences are different. On the other hand for periodic orbits with repetitions some of the above sequences coincide and the total number of associated sequences is a fraction of $n$. Let us  denote  $\Cl_{\bm n}$ the set of  all sequences of $n$ symbols corresponding to a cluster $\C_{\bm n}$ of periodic orbits.  In analogy with $\Z_2(n)$, $\Z_1(n)$  we define the following functions:
 \begin{equation}
 Z_2(n)=\sum_{ |\bm n|=n} |\Cl_{\bm n}|^2, \qquad Z_1(n)=\sum_{ |\bm n|=n} |\Cl_{\bm n}|,
\end{equation}
where $|\Cl_{\bm n}|$ is the number of sequences in $\Cl_{\bm n}$. For prime periodic orbits we just have $|\Cl_{\bm n}|=n|\C_{\bm n}|$. Since  vast majority of  periodic orbits are prime\footnote{In particular, if $n$ is a prime number there are only two non-prime periodic orbits. They correspond to sequences of all zeroes or ones, respectively}, see e.g., \cite{haake},  the connection between two functions is given by \cite{go1}:
\begin{equation}
 \Z_2(n)=\frac{1}{n^2} Z_2(n)(1+O(n^{-1})), \qquad \Z_1(n)=\frac{1}{n}Z_1(n)(1+O(n^{-1})). \label{connection}
\end{equation}

\subsection{Spectral problem}
To find the  size $|\Cl_{\bm n}|$  of $\bm n$'th cluster we need to count the number of closed paths which go through the edges $a\in X_p$ of  $G_p$  exactly $n_a$ times. To this end we introduce the connectivity matrix $Q$ between edges of the graph  and the auxiliary diagonal matrix $\Lambda(\bm \phi )$:

\begin{equation}
Q = 
\begin{pmatrix}
1&0 & \dots&0&1&0 & \dots&0\\
1&0 &\dots&0 &1&0 &\dots&0 \\
0&1 &\dots&0 &0&1 &\dots&0\\
0&1 & \dots&0 &0&1 &\dots&0\\
\vdots& \vdots& \ddots &\vdots&\vdots& \vdots& \ddots &\vdots\\
0&0 & \dots&1 & 0&0 & \dots&1\\
0&0 & \dots &1 & 0&0 & \dots &1
\end{pmatrix},
\quad
\Lambda(\bm \phi )=
\begin{pmatrix}
e^{i\phi_1}&0 &0& \dots&0 \\
0&e^{i\phi_2} &0&\dots&0  \\
\vdots& \vdots&  \ddots &\vdots& \vdots\\
0&0  &\dots&e^{i\phi_{2^p-1}} &  0\\
0&0  &\dots &0  & e^{i\phi_{2^p}}
\end{pmatrix}.
\end{equation}
The dimensions of these matrices are $2N\times2N $, with  $N:=2^{p-1}$.

The usefulness of introduction of matrices $Q$ and $ \Lambda(\bm \phi )$ is seen from the relation connecting traces of their powers with the sizes of the clusters $\Cl_{\bm n}$:
\begin{equation}
 \trace (Q \Lambda(\bm \phi ))^n=\sum_{\bm n} |\Cl_{\bm n}|\exp{\left( i(\bm n, \bm \phi )\right)}, \qquad  (\bm n, \bm \phi )=\sum_{a\in X_p}  n_a \phi_a,\quad n=\sum_{a\in X_p}  n_a,\label{KeyFormula}
\end{equation}
where the first sum runs  over all clusters $\Cl_{\bm n}$.  Eq.~(\ref{KeyFormula})  allows to express the second moment $Z_2$ as
\begin{equation}
Z_2=\sum_{\bm n} |\Cl_{\bm n}|^2=<|\trace (Q \Lambda(\bm \phi ))^n|^2>_{\bm \phi},\label{SecondMoment}
\end{equation}
with  the average  taken over the flat (Lebesgue) measure:
\begin{equation*}
<F(\bm \phi)>_{\bm \phi}=\prod_{a\in X_p}\int_{0}^{2\pi}\frac{d\phi_a}{2\pi} \,F(\bm \phi ). 
\end{equation*}
To simplify notation we denote the product of $Q$ and $\Lambda(\bm \phi )$ by
\[\Q(\bm \phi ):=\frac{1}{2}Q \Lambda(\bm \phi ).\]
Since half of the eigenvalues of $Q$ are zeroes \cite{GO},  the matrix $\Q(\bm \phi )$ has  $N$
zero eigenvalues for a general $\bm \phi$,  with the remaining $N$  distributed between $0$ and $1$. By
 eq.~(\ref{SecondMoment})  the function $Z_2$ can be expressed through the form factor of the ensemble of matrices $\Q(\bm \phi )$:
\begin{equation}
Z_2=\frac{2^{2n}}{N}K(n,N), \qquad  K(n,N)=\frac{1}{N}<|\sum_{i=1}^{N}\lambda^n_i(\bm \phi)|^2>_{\bm \phi},\label{SecondMoment1}
\end{equation}
where $\lambda_{i}(\bm \phi)$, $i=1,\dots N$ are non-trivial eigenvalues of  $\Q(\bm \phi )$.

The representation (\ref{SecondMoment1}) is the key component of our analysis, as it allows to relate $Z_2$ to the information on eigenvalue distribution in the ensemble of matrices $\Q(\bm \phi )$. For forthcoming treatment it is convenient to split 
$K(n,N)$ into diagonal and non-diagonal parts:
\begin{equation}
K^{(d)}(n,N)=\frac{1}{N}<\sum_{i=1}^{N}|\lambda_i(\bm \phi)|^{2n}>_{\bm \phi},\quad K^{(nd)}(n,N)=\frac{1}{N}<\sum_{i\neq j}^{N}\lambda^n_i(\bm \phi)\lambda^{n*}_j(\bm \phi)>_{\bm \phi}. \label{SecondMomentSplit2}
\end{equation}
The diagonal part is determined solely by the density  of the eigenvalues:
\[\rho(z)=\frac{1}{N}<\sum_{i=1}^{N}\delta(z-\lambda_i(\bm \phi))>_{\bm \phi}. \] 
 Since the flat measure is invariant under the shift $\phi_i\to\phi_i+c$, $i=1,\dots 2N$,   $\rho(z)$ depends only on the modulus $x=|z|^2$. Taking this into account  the diagonal part of the form factor  can be written down as:
 \begin{equation}
K^{(d)}(n,N)=\pi\int_0^1\rho(x)x^ndx.
\label{diagonal}
\end{equation}
In contrast, the non-diagonal part depends on the correlations between eigenvalues.

 \section{ Ensembles  of truncated unitary matrices}

 It is a simple observation that   $\Q(\bm \phi )$ are spectrally equivalent to the matrices of the truncated unitary type. More specifically,  we can  find a matrix $U_0$, such that 

\begin{equation}
U_0 \Q(\bm \phi )U_0^{\dagger}= PU(\bm \phi ),  \qquad P=\diag{1,\dots ,1, 0,\dots, 0}\label{trancated}
\end{equation}
with $U(\bm \phi )$ being   a unitary matrix and $P$ is a diagonal matrix containing $N$ zeroes and $N$ ones.
Indeed, by using  the unitary matrices  
\begin{equation}
U_0 = \frac{1}{\sqrt{2}}
\begin{pmatrix}
1&1 & \dots & 0&0\\
-1&1 &\dots &0&0 \\
\vdots& \vdots& \ddots & \vdots &\vdots\\
0&0 & \dots  &  1&1\\
0&0 & \dots  & -1 &1
\end{pmatrix},\qquad
U_1=\begin{pmatrix}
     I_{{N}\times {N}}& I_{{N}\times {N}}\\
     -I_{{N}\times {N}}& I_{{N}\times {N}}
    \end{pmatrix},
\end{equation}
where $I_{{N}\times {N}}$ is the ${{N}\times {N}} $ unit matrix,  we get the above  representation:
 \begin{equation}
  U_0 \Q(\bm \phi )U^{\dagger}_0=PU(\bm \phi ), \qquad U(\bm \phi )=U_1\Lambda(\bm \phi )U^{\dagger}_0.
 \end{equation}

\begin{figure}[htb]
\begin{center}{
\includegraphics[height=4.1cm]{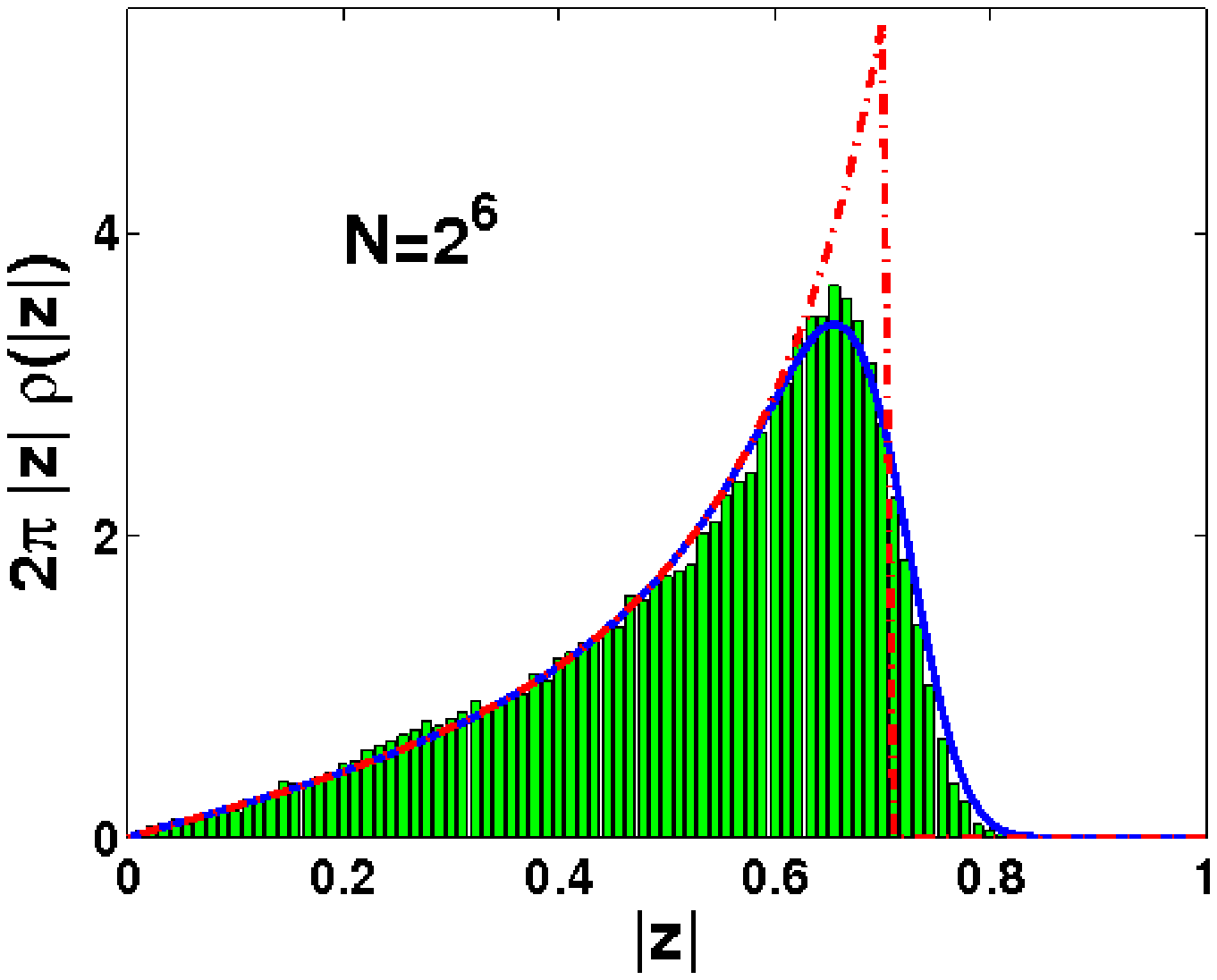}\hskip 0.1cm
\includegraphics[height=4.1cm]{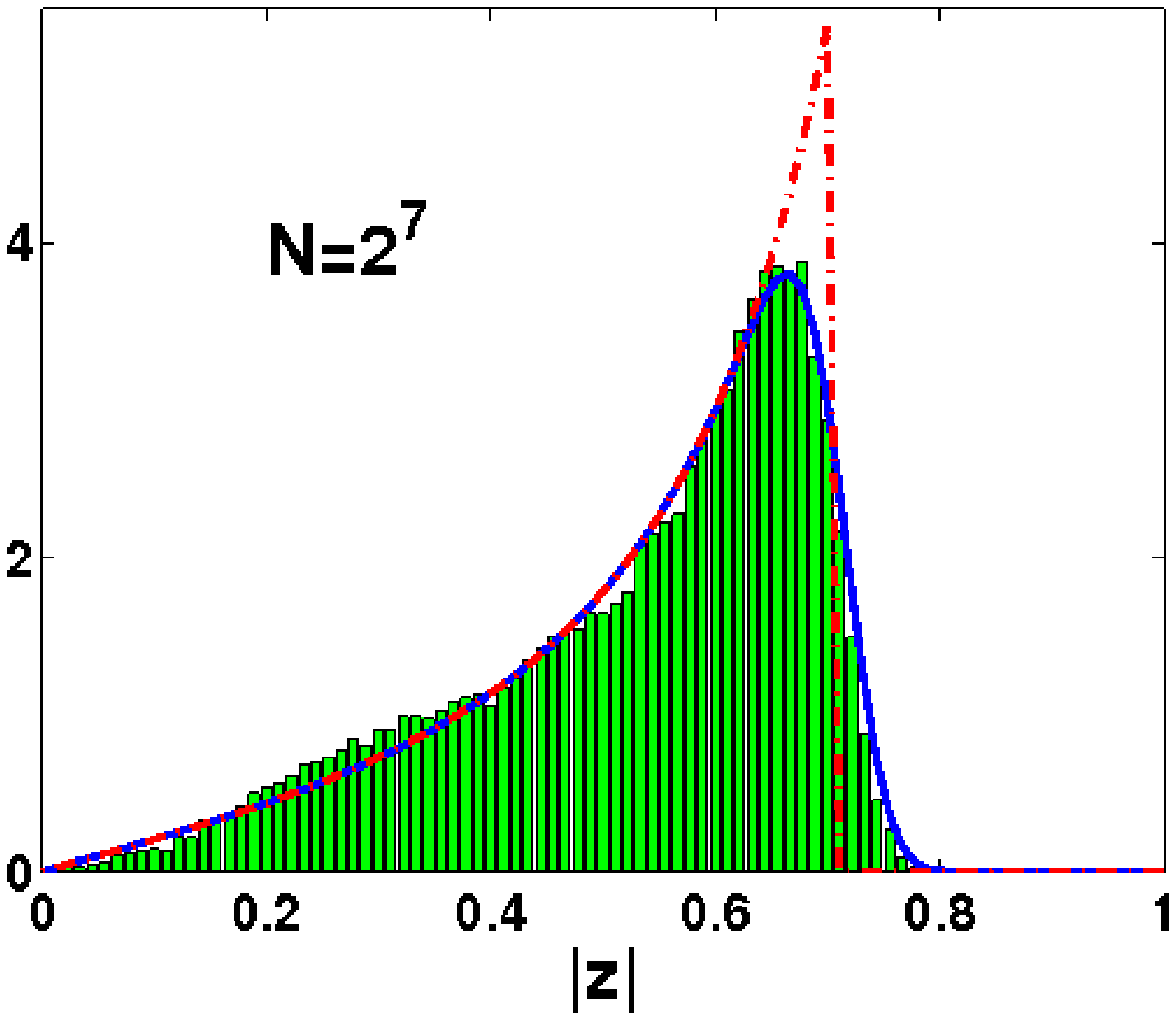}\hskip 0.1cm
\includegraphics[height=4.1cm]{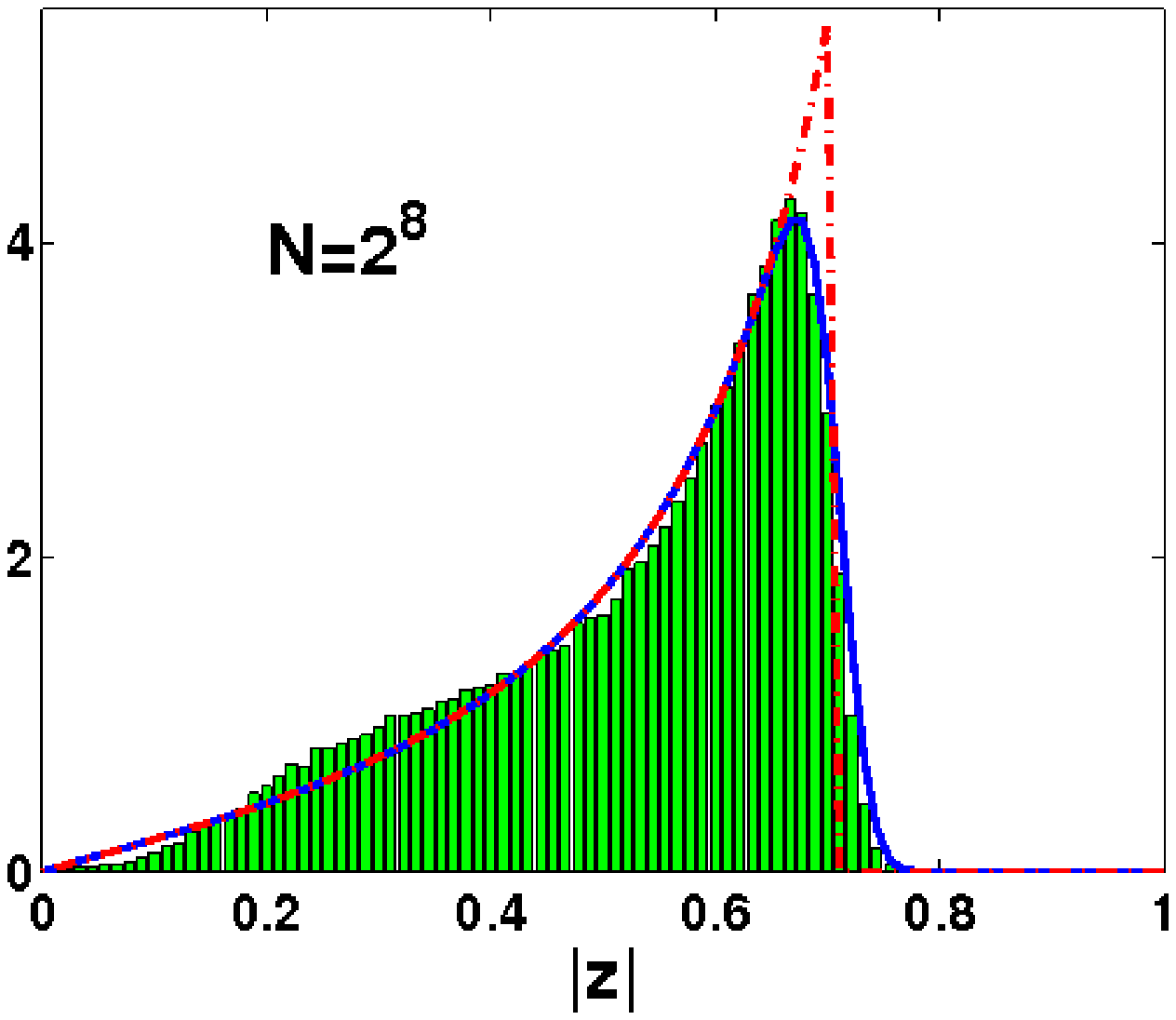}
}\end{center}
\caption{ \small{The spectral density for matrices $\Q(\bm \phi )$  in
comparison to the spectral density of truncated CUE (solid blue line). The
red (dashed) line shows limiting eigenvalue distribution  (\ref{limden}). Systematic deviations can be seen in the bulk of the spectrum.}
}\label{spectrdensity}
\end{figure}

The spectrum of matrices  $PU$, where $U$ is drawn from the circular unitary ensemble (CUE) has been previously  studied in \cite{som} for the projection $P$ of a general rank. In particular, if, as in our case, the rank of $P$ is half of its dimension   the   limiting $N\to\infty$  density of eigenvalues in these ensembles is given by:
\begin{equation}
\rho_{0}(x):=\lim_{N\to\infty}\rhot(x)=\frac{1}{\pi(1-x)^2} \qquad \mbox{ for $x\leq 1/2$   }, \label{limden}
\end{equation}
and $ \rho_{0}(x)=0$ otherwise, see fig.~\ref{spectrdensity}. The circle  $|z|=1/\sqrt{2}$, where the limiting density  abruptly goes to zero  is referred as the \textit{edge} of the spectrum.  For a finite $N$ the edge has a finite ``width`` of the order $1/\sqrt{N}$. The uniform asymptotics of spectral density at the vicinity of the edge  has been found in \cite{bog}. For the particular form of the projection matrix $P$  it reads:
\begin{equation}
  \rhot(x)=  \frac{1}{\pi(1-x)^2}\left(1-\frac{1}{2} \mbox{erfc}\left( \sqrt{4N}(1/2-x)\right) \right)+ O(N^{-\frac{1}{2}}),\label{main1}
\end{equation}
\begin{equation*}
 \mbox{erfc}(y)=\frac{2}{\sqrt{\pi}}\int^{\infty}_{y}e^{-x^2}dx. 
 \end{equation*}

The flat measure for the  ensemble of matrices $\Q(\bm \phi )$  obviously differs from the invariant measure of CUE. Nevertheless, the numerical simulation demonstrate   universality of spectrum properties at the vicinity of the edge. Namely, both eigenvalue density and the  eigenvalue correlations 
in ensemble of matrices  $\Q(\bm \phi )$ are the same as in the ensemble of truncated CUE on the scales $1/\sqrt{N}$ around the edge.  Below we summarize  these findings in the form of the conjecture:    

\begin{conj}
 1) Local density  $\rho(x)$ of eigenvalues $\lambda_i(\bm \phi)$  is universal on the scales $1/\sqrt{N}$ around $1/2$ i.e., for a fixed $s$ 
\begin{equation}
 \Delta\rho(s):=\rho\left(\frac{1}{2}-\frac{s}{\sqrt{N}}\right)-\rhot\left(\frac{1}{2}-\frac{s}{\sqrt{N}}\right)=O(N^{-\frac{1}{2}}),\label{Deltarho}
\end{equation}
where $\rhot(x)$ is the density of eigenvalues in the ensemble of truncated unitary matrices with the invariant measure, whose asymptotics is given by  (\ref{main1}).

2) Let $n=t\sqrt{N}$ and let 
 \[\Kk(t):=2^n\sqrt{N}K(t\sqrt{N},N),   \qquad \Kkt(t):= 2^n\sqrt{N}\Kt(t\sqrt{N},N)\]
 be rescaled form factors of respective ensembles of truncated unitary matrices.\footnote{The rescaling by factor $2^n$ is equivalent to the multiplication of each eigenvalue $\lambda_i(\bm \phi)$ by $\sqrt{2}$. Note that this shifts the edge of the spectrum of matrices $\Q(\bm \phi)$ to  $1$.}  In the limit, where $t$ is fixed and $N\to\infty$ both spectral form factors have the same asymptotics:
\begin{equation}
\Delta \Kk(t):=\Kk(t)-\Kkt(t) =O(N^{-\frac{1}{2}}).
\end{equation}
\label{conect}
\end{conj}

\begin{figure}[htb]
\begin{center}{
\includegraphics[height=4.7cm]{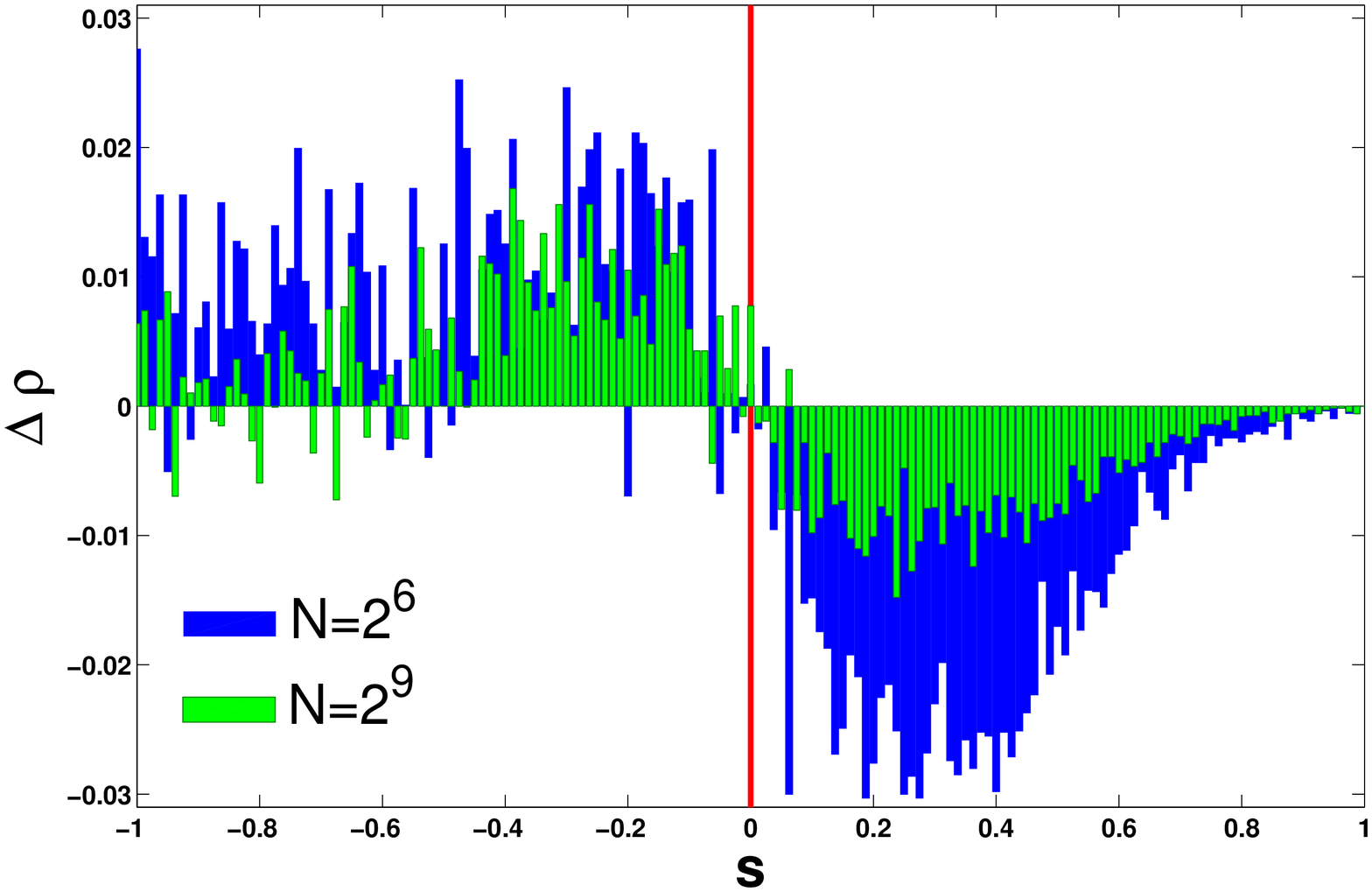}\hskip -0.2cm \includegraphics[height=4.7cm]{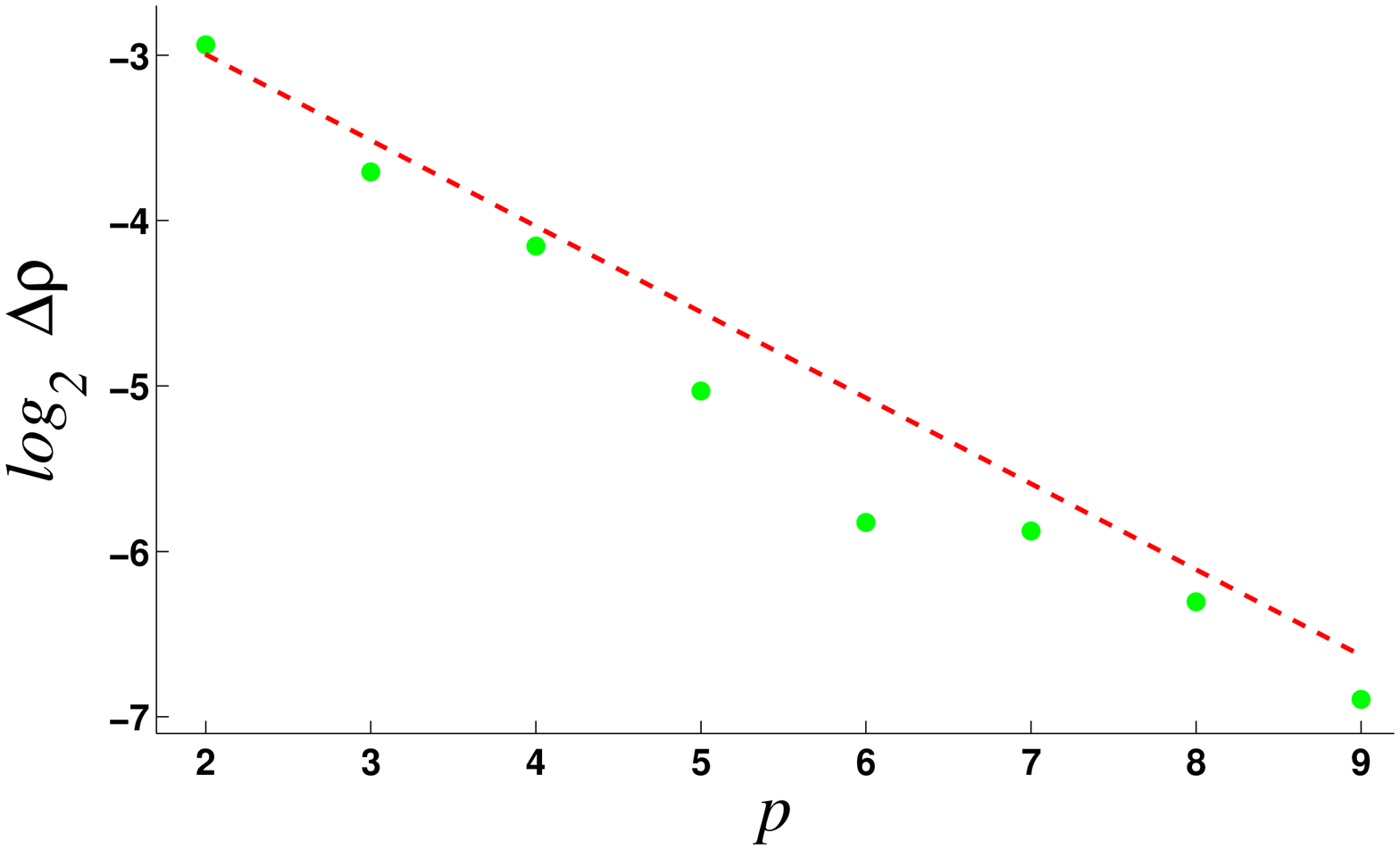}
}\end{center}
\caption{ \small{
Convergence of the radial eigenvalue density function of the matrix
ensemble $\Q(\bm\phi)$ to the density of the corresponding  truncated unitary
ensemble in the vicinity of the spectrum edge. The left plot shows
$\Delta\rho(s)$, see ~(\ref{Deltarho}), for two matrix sizes. On the
right picture the values of $\log_2 <\abs{\Delta\rho}>$ are plotted for
different $p$. The averaging is taken over the region $s\in(-1,1)$. 
The straight dashed (red) line with the slope $1/2$ is shown here for comparison.}  }\label{spectrdensity1}
\end{figure}

\begin{figure}[htb]
\begin{center}{
\includegraphics[height=3.7cm]{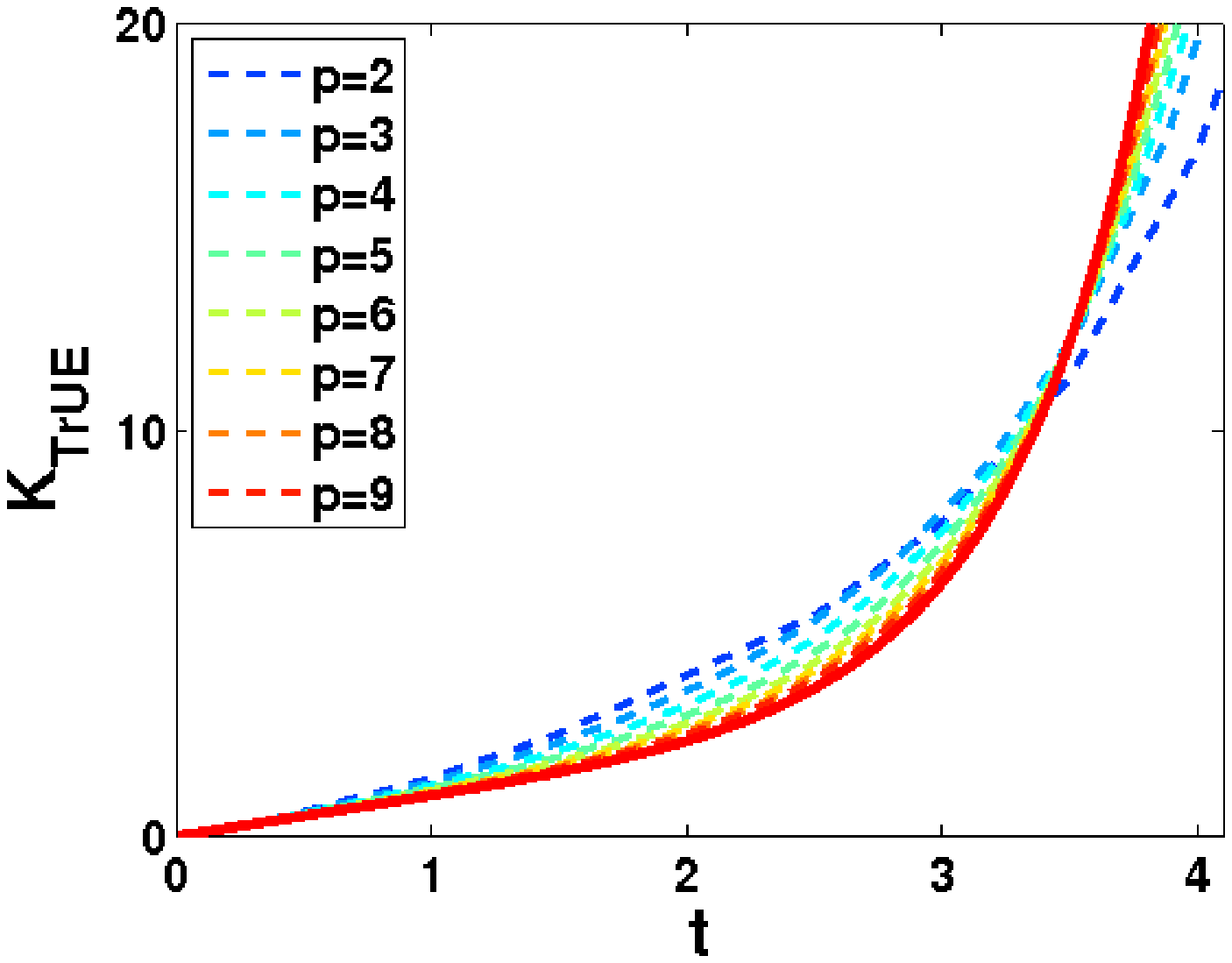} a)\hskip 0.2cm \includegraphics[height=3.7cm]{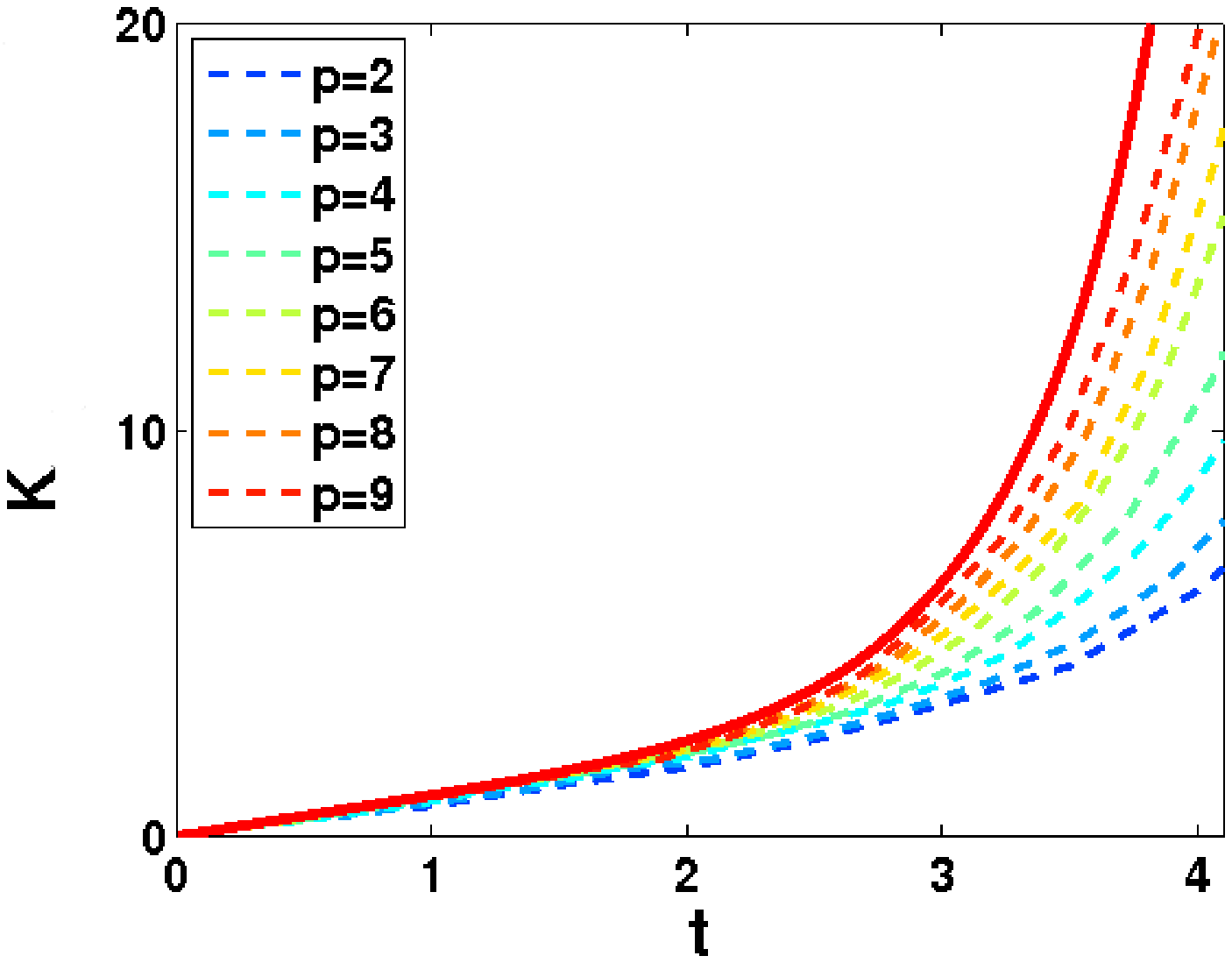}b)\hskip 4.2cm\includegraphics[height=3.7cm, width=4.7cm]{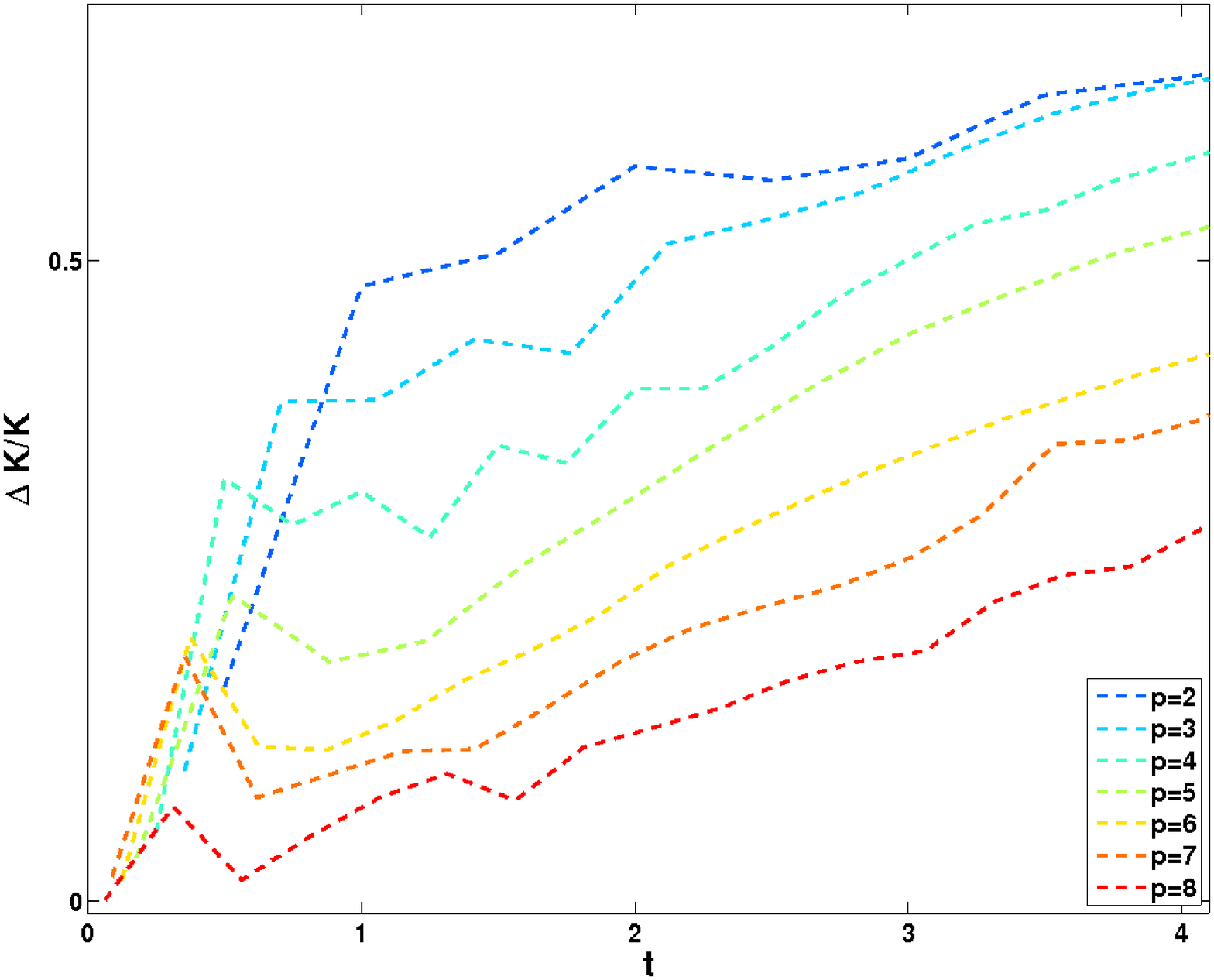} c)\hskip 0.2cm\includegraphics[height=3.7cm, width=4.7cm]{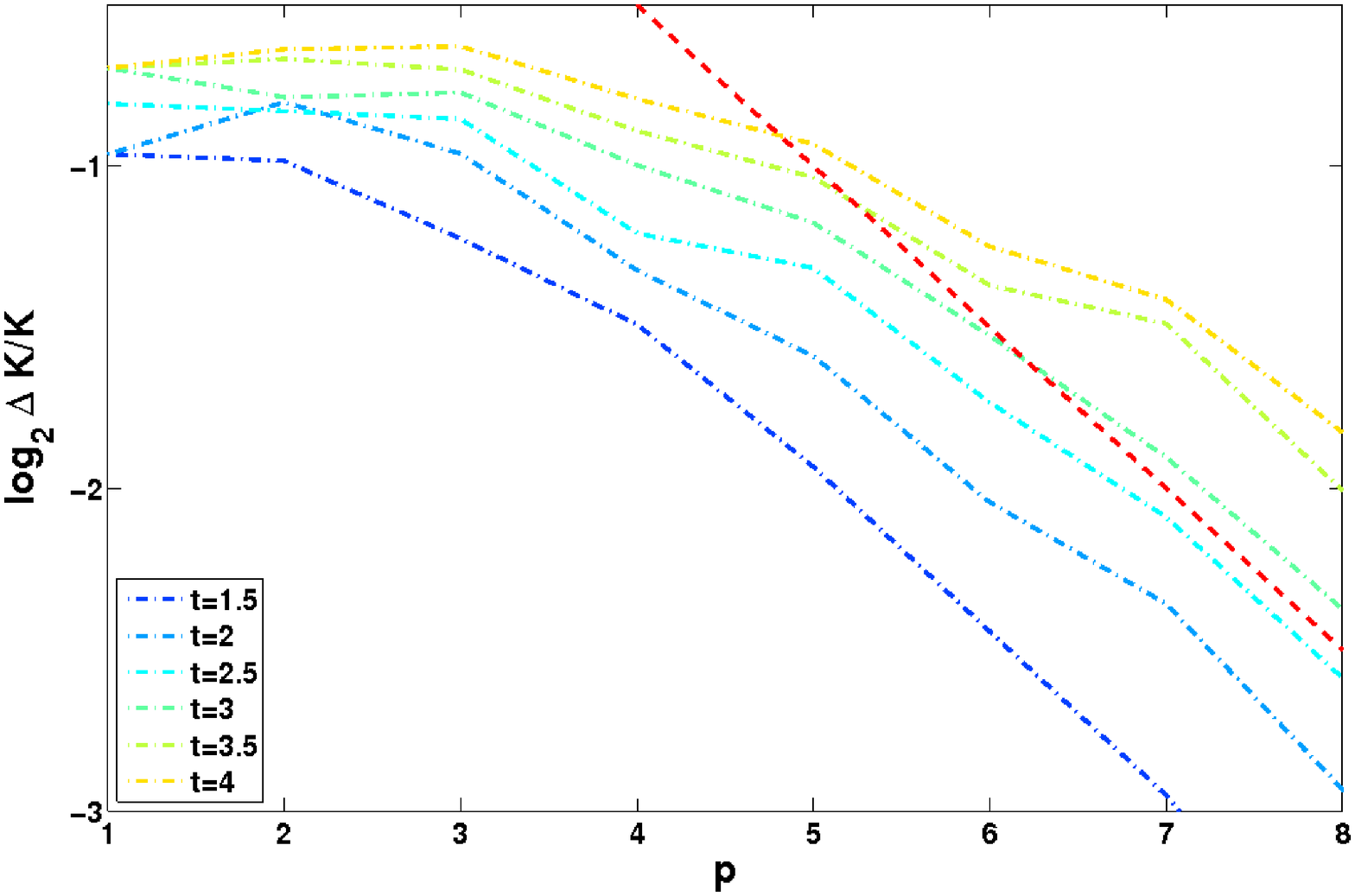} d)
}\end{center}
\caption{ \small{Rescaled spectral form factor $\Kkt(t)$ of  truncated CUE  is shown in the figure (a) as function of the parameter $t=n/\sqrt{N}$ for several parameters $p$ (dashed lines). In the figure (b) is similar plot for the rescaled spectral form factor  $ \Kk(t)$  of the ensemble of matrices $\Q(\bm \phi)$. The red solid lines  depicts asymptotics $\frac{4}{t}\sinh (t^2/4)$.  On the bottom figures  are the plots of the difference  $\Delta \Kk(t)$ as functions of $t$ (c) and  $p$ (d), respectively. The dashed straight line with the slope $1/2$ is depicted in figure (d) for comparison.  }  }\label{formfactor}
\end{figure}

To demonstrate the validity of this conjecture we have  numerically calculated spectrum of  matrices $\Q (\bm \phi )$ for thousands of  realizations of $\bm \phi$. The   averaged  over $\bm \phi$ form factor and density of states are shown  on figs.~\ref{spectrdensity1},\ref{formfactor}   for  different values of $N$. These plots  clearly demonstrate convergence of $\Kk(t)$ and $\rho(1/2-s/\sqrt{N})$ to their counterparts   in the corresponding ensembles of truncated CUE. 

\begin{rems}
1) Since the diagonal part of the form factor is determined by density $\rho$ only, the above conjecture actually implies that both 
$K^{(d)}$,  $K^{(nd)}$ asymptotically converge to their CUE counterparts.

 2) Recall that in the present work we consider symbolic dynamics arising in the baker's map.   For more complicated grammar rules the matrix $\Q(\bm \phi )$ is not  necessarily spectrally  equivalent to a matrix  of truncated unitary type. In general  this is a sub-unitary matrix which can be brought (by some unitary transformation) to  the form $DU$, where $D$ is a diagonal matrix with elements in the interval $[0,1]$ and $U$ is a unitary one. We believe that universality at the edge of the spectrum holds for these matrices 
as well.   Numerical simulations show that  spectral density and form factor asymptotics at the vicinity of the edge can be deduced by assuming that  $U$ belongs to CUE. The spectral density for  such ensembles has been studied  analytically  in \cite{fyod, bog}. 

3) So far  justification of the above conjecture  comes mainly from the numerics.  We believe, however, that analytical methods  based on the supersymmetry approach \cite{go2} can be used to prove the conjecture  and leave it for future investigation. 
\end{rems}

\section{ Uniform asymptotics of the form factor}

According to the above conjecture, in the regime where lengths of periodic  orbits are of the  order  $\sqrt{N}$ one can use the spectral  form factor of truncated CUE   in order to extract asymptotics of  $Z_2(n)$. 
 The goal of this section  is to derive the asymptotics of $\Kt(n)$ in the limit, where the ratio  $n/\sqrt{N}$  is  fixed.  

 It is instructive to separate  $\Kt(n,N)=\Kt^{d}(n,N)+\Kt^{nd}(n,N)$ into  diagonal and non-diagonal parts:
\begin{equation}
 \Kt^{(d)}(n)=\frac{1}{N}<\sum_{i=1}^{N} |z_i|^{2n}>, \qquad  \Kt^{(nd)}(n)=\frac{1}{N}<\sum_{i\neq j}^{N} z_i^{n*}z_j^{n}>,
\end{equation}
where we dropped label $N$ from the definition of $\Kt^{(d)}$, $\Kt^{(nd)}$ for the sake of notation compactness.
The exact analytical expressions for both parts can be easily obtained from the representation
\begin{equation}
 \Kt^{(d)}(n)=\frac{1}{N}\int d^2 z  R_1(z) |z|^{2n}, \qquad   \Kt^{(nd)}(n)=\frac{1}{N}\int d^2 z_{1} \int d^2 z_{2} R_2(z_{1},z_{2}) (z_{1},z^*_{2})^n,
\end{equation}
where $R_1$, $R_2$ are one-point  and two-point  spectral correlation functions, respectively.
Using the results of \cite{som} for  $R_1, R_2$  we obtain for the diagonal part: 
\begin{equation}
\Kt^{(d)}(n)=\frac{1}{N}\sum_{m=1}^{N}\frac{\Gamma(m+n)\Gamma(m+N)}{\Gamma(m)\Gamma(m+n+N)} \label{diagg}
\end{equation}
with $\Gamma(s)$ standing here for the gamma function. 
 The non-diagonal part is given by
\begin{equation} 
  \Kt^{(nd)}(n)=
-\frac{1}{N}\sum_{m=1}^{N-n}\frac{\Gamma(m+n)\Gamma(m+N)}{\Gamma(m)\Gamma(m+n+N)}\qquad  \mbox{ for   $n\leq N$}, \label{ndiagg}
\end{equation}
and $\Kt^{(nd)}(n)=0$ otherwise. To proceed further we use integral representation of gamma functions,
 \begin{align*} 
  &\frac{\Gamma(m+n)\Gamma(N)}{\Gamma(m+n+N)} =\int_{0}^{1}u^{m+n-1}(1-u)^{N-1}du, \\ &\frac{\Gamma(m+n)}{\Gamma(N)\Gamma(m)}=\frac{N}{2\pi}\int_{0}^{2\pi}(1+ e^{i\varphi })^{N+m-1}e^{-i\varphi N } d\varphi
 \end{align*}
to convert sums (\ref{diagg},\ref{ndiagg}) into double integrals: 
\begin{align}
\Kt^{(d)}(n)& =-\int_0^1\int_{0}^{2\pi}\frac{(1-x)^{N-1}x^{n+N}|1+e^{i\varphi }|^{2N} }{2\pi( 1-(1+ e^{i\varphi })x)} d\varphi\, dx+\pi\int_0^{1/2} \rho_0(x) x^n dx,
\\ 
\Kt^{(nd)}(n)& =\int_0^1\int_{0}^{2\pi}\frac{(1-x)^{N-1} x^N |1+e^{i\varphi }|^{2N} (1+e^{i\varphi })^{-n}}{2\pi(1-(1+ e^{i\varphi })x)} d\varphi \, dx-
\pi\int_0^{1/2} \rho_0(x) x^n dx,
\end{align}
where $\rho_0(x)$ is the limiting spectral density (\ref{limden}).
 We  will derive now the asymptotics of both diagonal and non-diagonal parts when  $n=t\sqrt{N}$, $N\to \infty $ and $t$ is fixed.

\subsection{Non-diagonal part} Let us first analyze the non-diagonal part. 
 After  change of the integration parameters $x\to 1/(x+1)$,  $e^{i\varphi }\to z$ the  integral (\ref{ndiagg}) can be cast into the  form:
\begin{multline}
\Kt^{(nd)}(n)=\int_0^{\infty}dx\oint_{\C} \frac {x^{N-1}(x+1)^{-2N}(1+z)^{2N-n}z^{-N-1}}{2\pi(x-z)}dz -
\pi\int_0^{1/2} \rho_0(x) x^n dx,\label{nondiag1}
\end{multline}
where the integration over  $z$  variable is performed along the circle $\C$ of the unit radius  centered at  the origin of the complex plane. The asymptotics of the first   integral $I$ in eq.~(\ref{nondiag1}) can be found using saddle point approximation. Writing dawn the above integral as:
\begin{gather}
 I=\int_0^{\infty}dx\oint_{\C}dz \frac {e^{NS(x,z)}}{2\pi(x-z)},\\  S=\frac{1}{N}\log\left(x^{N-1}(x+1)^{-2N}(1+z)^{2N-n}z^{-N-1}\right),
\end{gather}
one finds that the saddle point of $S$  is at $x=1, z =1+ O(1/\sqrt{N})$. Note that the denominator is of the order $1/\sqrt{N}$ at the saddle point.  We therefore make substitution $z=1+w$, $x=1+y$ and expand  $S$ up to the quadratic order in $w,y$, keeping the denominator intact. This gives:
 \begin{equation}  
2^n I =\frac {1}{2\pi}\int_{-1}^{+\infty}e^{-N y^2/4}\oint_{\C'} \frac{e^{-nw/2+Nw^2/4}}{y-w}dwdy +O\left(\frac{1}{N}\right).
\end{equation}
After rescaling   $\sqrt{N}w\to iv$, $\sqrt{N}y\to u$ and deforming the contour of the integration  we have: 
\begin{equation} 
 2^n I =\frac{I'(t)}{\sqrt{N}} +O\left(\frac{1}{N}\right), \quad I'(t)=\frac {1}{2\pi}\int_{-\infty}^{+\infty}\int_{-\infty}^{+\infty}\frac{e^{-u^2/4-tiv/2-v^2/4}}{u-iv}dv du.
\end{equation}
The last integral can be easily evaluated  by switching to polar coordinates $v=r\cos\theta$, $u=r\sin\theta$: 
\begin{multline} 
I'(t)=\frac{1}{2\pi}\int_{0}^{+\infty}\int_{0}^{2\pi}e^{-r^2/4-irt\sin\theta/2+i\theta}dr d\theta=\\
=2\int_{0}^{+\infty}e^{-r^2/4}J_1(rt/2)dr=\frac{2}{t}(1-e^{-t^2/4}).
\end{multline}

It remains to find  asymptotics of the second term $II$ in  eq.~(\ref{nondiag1}). The main contribution here comes from the boundary $1/2$ of the integration. As a result, to the leading order of $N$ we have:
\begin{equation} 
2^nII=\frac{\rho_{1}(\frac{1}{2})}{2n}\left( 1 +O\left(\frac{1}{N}\right)\right) =\frac{2}{t\sqrt{N}}+O\left(\frac{1}{N}\right).
\end{equation}
Combining $I$ and $II$ we  obtain for the non-diagonal part:
\begin{equation} 
\Kkt^{(nd)}(t):=\sqrt{N} 2^n\Kt^{(nd)}(n)=-\frac{2}{t}e^{-t^2/4}+O\left(\frac{1}{\sqrt{N}}\right).\label{resndiag}
\end{equation}
 
\subsection{Diagonal part} The asymptotics of the diagonal part  can be obtained in a similar way. Alternatively, one can use uniform  asymptotics   (\ref{main1}) for the  density of states and plug it into eq.~(\ref{diagonal}). The resulting expression is again the exponential function but with the opposite sign:
\begin{equation} 
\Kkt^{(d)}(t):=\sqrt{N} 2^n\Kt^{(d)}(n)=\frac{2}{t}e^{t^2/4}+O\left(\frac{1}{\sqrt{N}}\right). \label{resdiag}
\end{equation}
Remarkably, the leading asymptotics of the diagonal and non-diagonal parts turn out to be connected by a simple relationship:
\begin{equation} 
i\Kkt^{(d)}(it)\sim\Kkt^{(nd)}(t).
 \end{equation}

\section{ Estimation of cluster sizes}
\begin{figure}[htb]
\begin{center}{
\includegraphics[height=5.8cm]{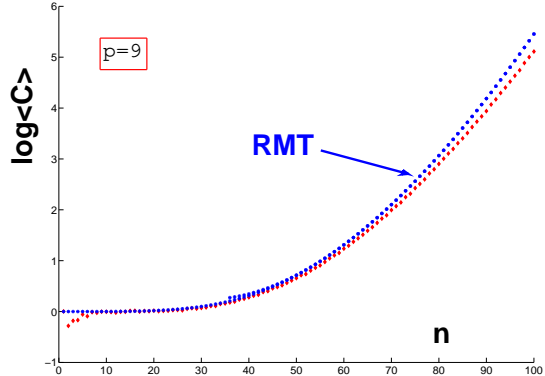}
}\end{center}
\caption{ \small{Comparison of RMT result (blue) (\ref{nice}) with the numerically calculated $\log<|C|>$ (red) as function of $n$ for   $p=9$. }  }\label{fig2}
\end{figure}

The conjecture (\ref{conect}) implies that for $n =t\sqrt{N} $ the asymptotics of  $K(n,N)$ can be extracted from eqs.~(\ref{resdiag}, \ref{resndiag}). Summing up the diagonal and non-diagonal parts of the form factor we obtain by eq.~(\ref{SecondMoment1}):
\begin{equation}
 \left(\frac{1}{2^n\sqrt{N}}\right) Z_2=\frac{4}{t}\sinh (t^2/4) + O(N^{-\frac{1}{2}}).\label{asympt}
\end{equation}
This is the main result of the present paper.
To better clarify  its meaning it is informative to  consider an average size of clusters rather than $Z_2(n)$ itself. Since  
 $Z_1(n)=2^n$,
by eq.~(\ref{connection}) the average size of clusters in the regime $n \sim\sqrt{N} $ can be estimated as
\begin{equation}
  <|\C|>=\frac{\sum_{|\bm n|=n} |\C_{\bm n}|^2}{\sum_{|\bm n|=n} |\C_{\bm n}|}=\frac{Z_2}{nZ_1}+O(N^{-\frac{1}{2}})=\frac{4}{t^2}\sinh (t^2/4) +O(N^{-\frac{1}{2}}),\label{nice}
\end{equation}
 see fig.~\ref{fig2}.
Let us show now  that  the parameter  $t^2=n^2/N$ actually  determines    the   number of encounters $\nu$  in  periodic orbits of the length $n$. To count the number of encounters  we will  follow the original approach of \cite{sr}. 

 A periodic sequence $x=\overline{x_1 x_2\dots x_n}$ is said to have the encounter $a$ of the length $s$ if the  subsequence $a=[a_1\dots a_s]$  appears (at least) twice in $x$ i.e.,:
\begin{align}
 x&=\overline{x_1\dots x_k\underbrace{a_1\dots a_s}_{encounter}x_{k+s+1}\dots x_m\underbrace{a_1\dots a_s}_{encounter}x_{m+s+1}\dots x_n}, \\
&\mbox{ where }  x_k\neq x_m, x_{k+s+1}\neq x_{m+s+1}.\nonumber
\end{align}
Let $[x], [y] $ be two (non-cyclic) sequences of the length $n$.  
We define the function $\delta^{(p)}$, such that $\delta^{(p)}([x],[y])=1$ if first $p $ symbols of  $[x]$, $[y] $ coincide and $\delta^{(p)}([x],[y])=0$, otherwise. The number  of encounters in a periodic sequence $x$ can be determined through the  function: 
\begin{equation} \N_x(p)=\sum_{i=1}^n\sum_{j}^n \delta^{p}(\sigma^j[x],\sigma^i[x]),\label{encountnum}\end{equation}
where $\sigma$ is  the shift map whose action on sequences reads as:
\[   \sigma\cdot[x_1 x_2\dots x_n]=[x_{n}x_1 \dots x_{n-1}].\]
Whenever the length $s$ of an encounter is larger than $p$ the function $\N_x(p)$ counts it  with the multiplicity  $2(s-p+1)$. 
One can, however, get read of  this redundancy by considering the difference:
\begin{equation}
 \nu(x)=(\N_x(p)-\N_x(p+1))/2,\label{encountnum1}
\end{equation}
where each encounter is counted only once.  Note that (\ref{encountnum1}), actually, counts  correctly only  encounters appearing twice in periodic sequences. This, however, does not affect the final result, since encounters of larger multiplicities  are rare in the considered regime. Indeed, for $n\ll N$ the probability  of getting encounter $k$-times scales as $n^2/N^k$.

We  can now   estimate $\N_x(p)$ applying  the uniformity principle.  According to it  long periodic orbits behave in the same way as trajectories with ``almost all'' initial conditions. Applying  the ergodicity theorem 
to the  $\N_x(p)/n^2$ and substituting the ``time'' average in (\ref{encountnum}) by the average over the phase space we obtain $\N_x(p)\sim n^2/N$ to the leading order in $n$. By eq.~(\ref{encountnum1}) this gives for the average number of encounters:
\begin{equation}
 \nu\sim n^2/4N, \qquad n\to\infty, \quad n\ll N.\label{encountnum2}
\end{equation}
Taking this into account we can finally express (\ref{verynice}) through the average number of encounters:
\begin{equation}
  <|\C|>=\frac{1}{\nu}\sinh \,\nu +O(N^{-\frac{1}{2}}).\label{verynice}
\end{equation}

\section{ Discussion}

It is well known that  spectral correlations in quantum chaotic systems essentially depend on considered time scale.  The spectral form factor of chaotic systems with broken time reversal symmetry has discontinuity in its first derivative at the Heisenberg time. From the semiclassical point of view such a jump  can be attributed to a  different behavior of periodic orbits for $n \ll N$ and $n \gg N$.
It is therefore interesting to see how this transition affects the average size $<|\C|>$ of clusters.

 Eq.~(\ref{verynice})  demonstrates that 
for  orbits with periods of the order of square root of the Heisenberg time, the asymptotics of $<|\C|>$  solely  depends on  the average number of encounters $\nu$. For $ n\lesssim \sqrt{N}$, when  encounters are rare,  we obtain  $\log<|\C|>\approx 0$.
 In other words, in this regime vast majority of  clusters  consist of just one periodic orbit  while the number of  clusters is equal to the number of periodic orbits.
As the average number of encounters $\nu$ becomes of the order one, an exponential growth of cluster sizes starts. For $\sqrt{N}\ll n\ll N$ we have  by (\ref{verynice})  $\log<|\C|> \sim n^2$, implying very fast growth of clusters, see fig.~\ref{fig2}.

The above result should be compared with one from \cite{go1} for the regime of very long trajectories $n\gg N$ -- far beyond the Heisenberg time. In that case the whole average (\ref{SecondMoment1}) is dominated by the largest eigenvalue $\lambda_{\max}(\phi)$ at the vicinity of   $\bm \phi =0$ leading to:   
\begin{equation}<|\C|>=\frac{2^{n}}{n}\left(\frac{N}{2n\pi}\right)^{N/4}  (1+O(1/n)).\label{goresult}\end{equation}
Here the average  size   of  clusters   grows with $n$ in the same way as the number of periodic orbits ($2^n/n$) while the number of clusters 
 grows only algebraically with $n$.
This is so, since at  $n\gtrsim N$ almost every point  of a generic periodic orbit belongs already to some  encounter. The further growth of  $n$ does not lead to  an essential increase in the number of encounters. This results in a  slower growth of cluster sizes:  $\log<|\C|> \sim n$. 

The transition between the aforementioned regimes is expected at $n\sim N$. At such scales only  the diagonal part of the form factor contributes to $Z_2(n)$. Furthermore,  the dominant contribution to the integral (\ref{diagonal}) comes from the interior of the interval  $[1/2,1]$, where the spectral density  is non-universal.  Consequently,  the results for  truncated CUE with the invariant measure are not applicable in this case. In general, one can assume that for $x>1/2$  the density is exponentially decaying  function of $N$:
\begin{equation}
 \rho(x)\sim\exp(-N\Phi(x)),
\end{equation}
where $\Phi(x)$ is a monotonic function of $x$ with $\Phi(1)=\infty$. Substituting this into eq.~(\ref{diagonal}) and performing saddle point approximation gives:
 \begin{equation}
<|\C|> \sim \exp n\left(-\frac{1}{\tau}\Phi(x_0)+\log(2x_0)\right),
  \end{equation}
where $n=\tau N$ and   $x_0$ is solution of the saddle point equation:
\begin{equation}
 x_0\Phi'(x_0)=\tau.
\end{equation}
Note that at the limit $\tau\gg 1$ (deep beyond the ``Heisenberg time'') $x_0\approx 1$ and   $<|\C|> \sim 2^n$ in agreement with 
(\ref{goresult}). For a more precise information on  the transition from the regime (\ref{verynice}) to (\ref{goresult}) one would need to know exact form of the function $\Phi(x)$ i.e., the non-universal tail of the spectral density.

In conclusion, let us mention few possible extensions of  the present paper results.   First,  general  symbolic dynamics with finite  grammar rules can be treated in a similar way. In this case cluster distribution can be obtained from the spectral form factor of   more general  ensembles of   sub-unitary matrices investigated in \cite{bog}, \cite{fyod}. We expect  that the final result (\ref{verynice}) stays the same, while connection between $\nu$ and $t$ depends on  symbolic dynamics. In particular, for trivial grammar rules with $l$ symbols this connection reads as $\nu\sim n^2(l-1)/2lN$ \cite{go2}.  Second, using our approach  systems with the time reversal symmetry can be treated, as well. In this case one should first redefine the notion of cluster.   If the time reversal symmetry is present, two periodic orbits belong to the same cluster iff they traverse  approximately the same points of the phase space up to a switch in the momentum direction. This additional freedom results in  a certain constrain on the elements of matrix $\Lambda(\bm \phi)$. Namely, for any  sequence $a=[a_1,\dots a_p]$ and its time reversal counterpart  $a^*=[a_p,\dots a_1]$ we have   $\phi_a=\phi_{a^*}$. In its own turn this  symmetry implies that effectively the matrix $\frac{1}{2}Q\Lambda(\bm \phi)$ belongs to the ensemble of truncated orthogonal, rather than general unitary matrices. Note that such ensembles with the Haar invariant measure were previously studied in \cite{som2}. We believe that these results can be  exploited in order to obtain information on the cluster distribution for periodic orbits in the dynamical system with time-reversal invariance.

\section*{Acknowledgments}
We thank S. Kumar  for valuable discussions and help with the derivation of eqs.~(\ref{diagg},\ref{ndiagg}). Financial support by the SFB/TR12 and Gu 1208/1-1 research grant of the Deutsche
Forschungsgemeinschaft is gratefully acknowledged.

\end{document}